\documentstyle[editedvolume,epsfig]{crckapb}

\begin{opening}

\title{COSMIC GAMMA-RAY BURSTS\protect\\}
\subtitle{The most energetic phenomenon in the Universe}

\author{A.J. CASTRO-TIRADO}
\institute{Laboratorio de Astrof\'{\i}sica Espacial\\
           y F\'{\i}sica Fundamental (LAEFF-INTA)\\
	   P.O. Box 50727, E-28080, Madrid, Spain}
\institute{Instituto de Astrof\'{\i}sica de Andaluc\'{\i}a (IAA-CSIC)\\
	   P.O. Box 03004, E-18080, Granada, Spain}

\end{opening}

\runningtitle{Cosmic gamma-ray bursts}

\begin{document}


\section{Abstract}

GRBs have remained a puzzle
for many high--energy astrophysicists since their discovery in 1967.
With the advent of the X--ray satellites {\it BeppoSAX} and {\it RossiXTE}, 
it has been possible to carry out deep multi-wavelength observations of 
the counterparts associated with the GRBs just within a few hours of 
occurence, thanks to the observation of the fading X-ray emission that 
follows the more energetic gamma-ray photons once the GRB event has ended. 
The fact that this emission (the afterglow) extends at longer wavelengths, 
has led to the discovery of the first optical/IR/radio counterparts in 
1997-98, greatly improving our understanding of these sources.
Now it is widely accepted that GRBs originate at cosmological
distances but while the observed afterglow satisfies the predictions of the 
relativistic fireball models, the central engines that power these 
extraordinary events remain still unknown. Detailed results for 
nine selected GRBs are presented as well as a summary of the GRB counterparts 
and host galaxies found so far.

\section{Introduction. General characteristics.}

In 1967-73, the four VELA spacecraft (named after the spanish verb 
{\it velar}, to keep watch), that where originally designed for verifying 
whether the former Soviet Union abided by the Limited Nuclear Test Ban 
Treaty of 1963, observed 16 peculiarly strong events 
(Klebesadel, Olson and Strong 1973, Bonnell and Klebesadel 1996).
On the basis of 
arrival time differences, it was determined that they were related neither
to the Earth nor to the Sun, but they were of cosmic origin. Therefore they
were named cosmic Gamma-Ray Bursts (GRBs hereafter).\par

GRBs appear as brief flashes of cosmic high energy photons, emitting the 
bulk of their energy above $\approx$ 0.1 MeV. 
They are detected
by instruments somewhat similar to those used by the particle physicists at
their laboratories, but the main difference is that GRB detectors have to 
be placed onboard balloons, rockets or satellites.

The KONUS instrument on {\it Veneras 11} and {\it 12} gave the first 
indication that GRB sources were isotropically distributed in the sky 
(Mazets et al. 1981, Atteia et al. 1987).  Based on a much larger sample, this 
result was nicely confirmed by BATSE on board the {\it CGRO} satellite, 
launched in the spring of 1991, an instrument that has revolutioned the GRB 
field (Meegan et al. 1992). Since then, BATSE is detecting about 300 GRBs 
every year, but only very few can be localized accurately. The apparent 
isotropy of the bursts in the sky 
ruled out the models dealing with neutron stars in the Galactic Plane, and 
it was rather interpreted in terms of GRBs arising at cosmological distances, 
although the possibility of a small fraction of the sources lying nearby, 
within a galactic disc scale of few hundred pc, or in the halo of the Galaxy, 
could not be discarded by that time. 
Another result was that the time profiles of the bursts
are very different, with some GRBs lasting a few ms and others lasting for 
several minutes.  In general, there was no evidence of periodicity in the 
time histories of GRBs.  However there was indication of a bimodal 
distribution of burst durations, with
$\sim$25\% of bursts having durations around 0.2 s and $\sim$75\% with 
durations around 30 s. 
A deficiency of weak events was noticed in the log $N$-log $S$ diagram,
as the GRB distribution deviates from the -3/2 slope of the straight line 
expected
for an homogeneous distribution of sources assuming an Euclidean geometry. 
All these observational data led many researchers to believe that GRBs are 
indeed at cosmological distances. However, the GRB distance scale had to 
remain unknown for 30 years.  A comprenhensive review of these observational 
characteristics can be seen in Fishman and Meegan (1995).

\section{The search and detection of counterparts at other wavelengths}

It was well known that an important clue for solving the GRB puzzle was going
to be the detection of transient emission -at longer wavelengths- associated 
with the bursts.
A review on the unsuccessful search for counterparts prior to 1997 
can be seen in Castro-Tirado (1998) and references therein.
Here I will summarize some results concerning nine selected bursts 
detected by the {\it BeppoSAX} ({\it BSAX}) and {\it RossiXTE} 
({\it RXTE}) satellites in 1996-98. See also Piran (1999).


\subsection{GRB 960720}
This was the first GRB for which a precise position (5$^{\prime}$ error radius)
was obtained by {\it BSAX} (in\' \rm t Zand et al. 1997). The bright 
radio-loud quasar 4C 49.29 at $z$ = 1.038 lies within the tiny GRB error box. 
The probability to find such object within the error box is 
2 $\times$ 10$^{-4}$ (Greiner and Heise 1997, Piro et al. 1997a), 
but no firm relationship could be established.  

\subsection{GRB 970111}
The position for this GRB was rapidly distributed, 
allowing to perform deep optical imaging only 19 hr after the high
energy event. No optical variability was found within the error box
down to B = 23, R = 22.5 (Castro-Tirado et al. 1997, Gorosabel et al.
1998a). A careful analysis of the X-ray data revealed a weak 
fading X-ray source within the GRB error box (Feroci et al. 1998a), 
but its association to GRB 970111 could not be definetively proven.  

\subsection{GRB 970228}
Thanks to {\it BSAX}, it was possible on 28 Feb 
1997 to detect the first {\it clear} evidence of a long X-ray tail 
 -the X-ray afterglow- following GRB 970228. A previously unknown 
X-ray source was seen to vary by a factor of 20 on a 3 days timescale. 
The X-ray fluence was $\sim$ 40 \% of the gamma-ray fluence, as reported 
by Costa et al. (1997b), implying that the X-ray 
afterglow was not only the low-energy tail of the GRB, but also a significant 
channel of energy dissipation of the event on a completely different timescale.
Another important result was the non-thermal origin of the burst radiation 
and of the X-ray afterglow (Frontera et al. 1998a). The precise X-ray 
position (1$^{\prime}$) led to the discovery of the first optical transient 
(or optical afterglow, OA) associated to a GRB, identified on 28 Feb 1997, 
20 hr after the event, by Groot et al. (1997).
The OA was afterwards found on earlier images taken by Pedichini et al. 
(1997) and Guarnieri et al. (1997), 
in the rising phase of the light curve. The maximum was reached $\sim$ 20 hr 
after the event (V $\sim$ 21.3), and followed by a power-law decay  
F $\propto$ t$^{-1.2}$ (Galama et al. 1997, Bartolini et al. 1998). 
An extended source was seen at the OA position since the very beginning 
by both ground-based and {\it HST} observations (van Paradijs et 1997, 
Sahu et al. 1997).
New {\it HST} observations taken 6 months after the event were reported by 
Fruchter et al. (1997). Both the OA (at V = 28) and the extended source 
(V = 25.6) were seen. 
The extended object surrounding the point-source was  interpreted as a 
galaxy, according to the similarities (apparent size, magnitude) with 
objects in the {\it HST} Deep Field. Despite exhaustive efforts in order
to get spectroscopic measurements, no emission lines were
found, implying that the redshift of this galaxy should lie in the range
1.3 $\leq$ $z$ $\leq$ 2.5 (Fruchter et al. 1999). 

  \begin{figure}[t]
    \centering
     \epsfig{file=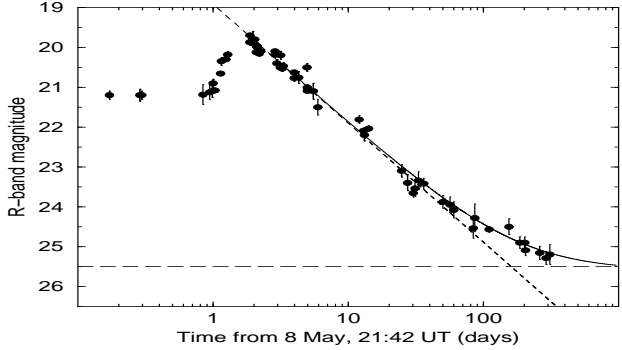,width=13.5cm, height=5cm, angle=0.0}
  \caption{The R-band light curve of the GRB 970508 optical afterglow, from
           data quoted in this paper. The dotted like is the contribution of 
           the GRB afterglow itself, following F $\propto$ t$^{-1.19}$ two 
           days after the burst. The horizontal dashed line is the R = 25.5 
           host galaxy, whereas the solid line is the contribution of both 
           (afterglow plus host galaxy).}
   \end{figure}



\subsection{GRB 970508}
The second OA associated to a GRB was discovered by Bond (1997) 
within the GRB 970508 error box, and observed 3 hr after the burst 
in unfiltered images (Pedersen et al. 1998). The optical light 
curve reached a peak in two days (R = 19.7, Castro-Tirado et al. 1998a, 
Djorgovski et al. 1997, Galama et al. 1998a) and 
was followed by a power-law decay F $\propto$ t$^{-1.2}$. 
Optical spectroscopy allowed a 
direct determination of a lower limit for the redshift of GRB 970805 
($z \geq 0.835$) and was the first proof that at least a fraction of the GRB
sources lie at cosmological distances (Metzger et al. 1997). The 
flattening of the decay in late August 1997 (Pedersen et al. 1998, Sokolov et
al. 1998a) revealed the contribution of a constant brightness source 
-the host galaxy- that was revealed in late-time imaging obtained
in 1998 (Bloom et al. 1998a, Castro-Tirado et al. 1998b, 
Zharikov et al. 1998). See Fig. 1. 
The luminosity of the galaxy is well below the knee of the galaxy luminosity 
function, L $\approx$ 0.12 $L^{*}$, and the detection of deep 
Mg I absorption (during the bursting episode) and strong [O II] 3727 $\rm \AA$ 
emission (the latter mainly arising in H II regions within the host galaxy) 
confirmed $z$ = 0.835 and suggested that the host could be a normal dwarf 
galaxy (Pian et al. 1998a), with a star formation 
rate (SFR) of $\sim$ 1.0 $M_{\odot}$ year$^{-1}$ (Bloom et al. 1998a).
Prompt VLA observations of the GRB 
970508 error box allowed detection of a variable radio source at 1.4, 4.8 
and 8.4 GHz, the first radiocounterpart ever found for a GRB (Frail et al. 
1997). The fluctuations could be the result of 
strong scattering by the irregularities in the ionized Galactic 
interstellar gas, with the damping of the fluctuations with time indicating 
that the source expanded to a significantly larger size.
However VLBI observations did not resolve the object (Taylor et al. 1997).
The transient was also detected at
15 GHz (Pooley and Green 1997) and as a continuum point source at 86 GHz with 
the IRAM PdBI on 19-21 May 1997 (Bremer et al. 1998). 
A Fe K$\alpha$ line redshifted at $z$ = 0.835 in the X-ray 
afterglow spectrum was reported (Piro et al. 1999), and could 
be attributed to a thick torus surrouding the central engine 
(M\'esz\'aros and Rees 1998).

\subsection{GRB 970828}
This burst was detected by {\it RXTE} (Remillard et al. 1997) and was followed
up by {\it ASCA} and {\it ROSAT} (Murakami et al. 1997, Greiner et al. 1997).
The X-ray spectrum as seen by {\it ASCA} is strongly absorbed 
(Yoshida et al. 1999), suggesting that the event occurred in a dense medium. 
The fact that no optical counterpart down to R = 23.8 was detected between 
4 hr and 8 days after the event, could support the idea that the non-detection
was due to photoelectric absorption (Groot et al. 1998). An excess at 6.7
keV was foud by {\it ASCA} in the X-ray afterglow spectrum. If this is due
to highly ionized Fe, then $z$ $\sim$ 0.33 (Yoshida et al. 1999). 

\subsection{GRB 971214}
The third optical transient was related to GRB 971214 and identified
by Halpern et al. (1997) as an I = 21.2 object that faded 1.4 mag in 1
day. Independently, this source was also noticed as suspicious by Itoh et 
al. (1997). Further observations proved that the decay followed a power-law 
decline (Diercks et al. 1997, Castander et al. 1997) with 
F $\propto$ t$^{-1.4}$, 
similar to, but steeper than, the previous GRBs with optical counterparts. 
GRB 971214 was detected in the K-band 4 hr after the burst (Gorosabel et al. 
1998b) and rapidly decreasing in the J-band (Tanvir et al. 1997).  
However, no source was detected at 850 $\mu$m at the
James Clerk Maxwell Telescope. The upper limit was 1 mJy on Dec 17-22 
(Smith et al. 1999).
Spectroscopy of the host galaxy (R = 25.6) one month later reveled a 
strong emission-line attributted to Ly-$\alpha$ redshifted at $z$ = 3.42
(Kulkarni et al. 1998a), implying a SFR of
 1.0 $\pm$ 0.5 $M_{\odot}$ year$^{-1}$, similarly to other galaxies at
comparable redshift. The emitted energy, 
assuming isotropic emission, was 3 $\times$ 10$^{53}$ erg.

\subsection{GRB 980329}
Together with GRB 970111, this burst is among the top 2$\%$ burst with 
larger gamma-ray fluences as detected by BATSE. The detection of a
variable radiosource (Taylor et al. 1998) within the GRB error box
led to the indentification of an optical transient (Klose et al. 1998,
Palazzi et al. 1998). The evidence for a dusty host
(R-K = 4.7) was confirmed by a SCUBA detection at 850 $\mu$m (Smith et al.
1999) and Fruchter (1999) proposed that the host could lie
at $z$ $\sim$ 5. In that case, if the gamma-rays were radiated 
isotropically, the implied energy would be 5 $\times$ 10$^{54}$ erg ! 

\subsection{GRB 980425}
A peculiar supernova (SN 1998bw) has been found in the WFC error box for 
this soft GRB (Galama et al. 1998b). The SN lies in the galaxy ESO 184-82 (at 
$z$ = 0.0085). The fact that the SN event occurred within $\pm$ 1 day 
of the GRB event, together with the relativistic expansion speed derived from 
the radio observation (Kulkarni et al. 1998b) strengths such a relationship. 
In that case, the total energy released would be 8 $\times$ 10$^{47}$ erg.
However, the fact that a fading X-ray source -as in {\it all} the previous
cases- unrelated to the SN was detected by {\it BSAX} in the 
GRB error box (Pian et al. 1998b, Piro et al. 1998) 
cast some doubts on the SN/GRB association (Graziani et al. 1999).

\subsection{GRB 980703}
A variable radiosource was found by Frail et al. (1998), and the optical
counterpart was independently discovered by Frail et al. (1998) and 
Zapatero-Osorio et al. (1998) in the error box provided by {\it RXTE}.  
The host galaxy is the
brightest one so far detected (R = 22.5, H = 20.5) according to Bloom et
al. (1998b) and Castro-Tirado et al. (1999). Optical spectroscopy
revealed the [O II] emission line at $z$ = 0.966, as well as some Fe II y 
Mg II absorption ones. The derived SFR is 
$\sim$ 63 $M_{\odot}$ year$^{-1}$ (Djorgovski et al. 1998). 
The released energy during the GRB event amounts to 
$\sim$3 $\times$ 10$^{52}$ erg. 

\vspace{0.5cm}
Further X-ray afterglows were observed by {\it BSAX} and {\it RXTE} in 
1998 (GRB 980706, GRB 981220 and GRB 981226). Exponents for the 
power-law decay in the X-rays and in the optical are in the range
$\alpha$ = 1.10-2.25 for a dozen of bursts. These results are given 
on Table 1 whereas Table 2 summarizes the properties of the host galaxies 
found so far. See also (Greiner 1999) for an updated information.

\begin{table}[bt]
\begin{center}
\caption{GRBs detected by {\it BeppoSAX} and {\it RXTE} in 1996-98}
\begin{tabular}{lllll}
\hline
GRB    & X-rays   &optical-IR& radio & References (X-ray detection)\\
\hline
960720 &          &          &       &  Piro et al. (1996)\\
970111 &  yes ?   &   no     &  no   &  Costa et al. (1997a)\\
970228 &  yes     &   yes    &  no   &  Costa et al. (1997b)\\
970402 &  yes     &   no     &       &  Piro et al. (1997b)\\
970508 &  yes     &   yes    &  yes  &  Piro et al. (1997c)\\
970616 &  yes ?   &   no     &  no   &  Marshall et al. (1997)\\
970815 &  yes ?   &   no     &  no   &  Smith et al. (1997)\\
970828 &  yes     &   no     &       &  Remillard et al. (1997a)\\
971214 &  yes     &   yes    &  no   &  Antonelli et al. (1997)\\
971227 &  yes ?   &   yes ?  &       &  Sofitta et al. (1997)   \\
980109 &          &   yes ?  &       &  in\' \rm t Zand et al. (1998)\\
980326 &  yes     &   yes    &       &  Celidonio et al. (1998) \\
980329 &  yes     &   yes    &  yes  &  Frontera et al. (1998b) \\
980425 &  yes ?   &   yes ?  & yes ? &  Sofitta et al. (1998)   \\
980515 &  yes ?   &          &       &  Feroci et al. (1998b)   \\
980519 &  yes     &   yes    &       &  Muller et al. (1998)    \\
980613 &  yes     &   yes    &       &  Smith et al. (1998a)    \\
980703 &  yes     &   yes    &  yes  &  Levine et al. (1998)    \\
980706 &  yes ?   &          &       &  Hurley et al. (1998)    \\
981220 &  yes     &          &       &  Smith et al. (1998b)    \\
981226 &  yes     &   yes ?  & yes ? &  Di Ciolo et al. (1998)  \\
\hline
\end{tabular} 
\end{center}
\end{table}

\begin{table}[bt]
\begin{center}
\caption{GRB host galaxies}
\begin{tabular}{llcll}
\hline
GRB    & R (host) &       $z$    &SFR ($M_{\odot}$ year$^{-1}$)&  References\\
\hline
970228 &  25.2    & 1.3$\leq z \leq$2.5  &       &  Fruchter et al. (1998)  \\
980828 & $\geq$24 &         0.33 ?       &       &  Yoshida et al. (1999)   \\
970508 &  25.7    &         0.835        &   1   &  Bloom et al. (1998a)    \\
971214 &  25.6    &         3.418        &   5   &  Kulkarni et al. (1998a) \\
980326 &$\geq$27.3&                      &       &  Bloom and Kulkarni (1998)\\
980329 &$\geq$25.5&        $\sim$5 ?     &       &  Fruchter (1999)  \\
980519 & $\sim$26 &                      &       &  Hjorth et al. (1999)    \\
980613 &  23.8    &         1.096        &   3   &  Djorgovski et al. (1999) \\
980703 &  22.5    &         0.966        &  63   &  Djorgovski et al. (1998)\\
\hline
\end{tabular} 
\end{center}
\end{table}


\section{Theoretical models}

The observational characteristics of the GRB counterparts can be
accommodated in the framework of the relativistic fireball models, first 
proposed by Goodman (1986) and Paczy\'nski (1986), in which a compact source 
releases 10$^{53}$ ergs of energy within dozens of seconds. See also  
M\'esz\'aros et al. (1994), M\'esz\'aros and Rees (1997), Tavani (1997), 
Vietri (1997), Waxman (1997) and Wijers et al. (1997).  
The opaque radiation-plasma accelerates
to relativistic velocities (the fireball). The blast wave is moving ahead
of the fireball, and sweeps up the interstellar matter, producing an afterglow
at frequencies gradually declining from X-rays to visible and radio 
wavelenghts. An extensive discussion is given by Piran (1999).

The properties of the blast wave can be derived from the classical synchrotron
spectrum (Ginzburg and Syrovatskii 1965) produced by a population of 
electrons with the addition of self absorption at low frequencies and 
a cooling break (Sari, Piran and Narayan 1998). Thus, the four 
main quantities involved are: 

i) the synchrotron frequency: the population of electrons has a power-law 
distribution of Lorentz factors $\Gamma_{e}$ following d$N$/d$\Gamma_{e}$ 
$\propto$ $\Gamma_{e}^{-p}$ above a minimum Lorentz factor
$\Gamma_{e}$ $\geq$ $\Gamma_{m}$, corresponding to the 
synchrotron frequency $\nu_{m}$; 

ii) the break frequency: if the electrons are energetic they will cool 
rapidly for a break frequency $\nu_{c}$ $<$ $\nu_{m}$, 
and low energy electrons will have always slow cooling 
for $\nu_{c}$ $>$ $\nu_{m}$, 

iii) the self-absorption frequency: synchrotron self absorption
becomes important below a critical frequency $\nu_{a}$, and

iv) the maximum flux F$_{\nu, max}$. 

After some time, the evolution of the blast wave is adiabatic, and 
$\nu_{c}$ $>$ $\nu_{m}$ implying that
F$_{\nu}$ $\propto$ $\nu^{-(p-1)/2}$ 
for $\nu_{m}$ $\leq$ $\nu$ $<$ $\nu_{c}$; 
F$_{\nu}$ $\propto$ $\nu^{-p/2}$ for $\nu$ $>$ $\nu_{c}$
y F$_{\nu}$ $\propto$ $\nu^{1/3}$ for $\nu$ $<$ $\nu_{m}$.
For $\nu_{a}$ $\ll$ $\nu_{m}$, F$_{\nu}$ $\propto$ $\nu^{2}$.

The time evolution of the flux as a function of a given
frequency is a function of
$\nu_{c}$ ($\propto$ t$^{-1/2}$) and $\nu_{m}$ ($\propto$ t$^{-3/2}$).
For $\nu$ $\leq$ $\nu_{c}$ the flux decays following a power-law with 
exponent ${\alpha}$, F $\propto$ t$^{\alpha}$ with $\alpha$ = (2$-$3$p$)/4,
thus allowing to determine $p$.

Thus, the determination for every GRB of the six observables $\nu_{m}$, 
$\nu_{c}$, $\nu_{a}$, F$_{\nu, max}$, $p$ (from the multiwavelength 
spectrum, see Fig. 2) and $z$ (from optical or X-ray spectroscopy) allows 
to obtain the total energy per solid angle (E$_{52}$), 
the fraction of the shock energy in electrons ($\epsilon_{e}$), the
fraction of the shock energy in post-grb magnetic fields ($\epsilon_{B}$) 
and the density of the ambient medium ($n$). See for example, Wijers
and Galama (1998).

  \begin{figure}[t]
    \centering
    \epsfig{file=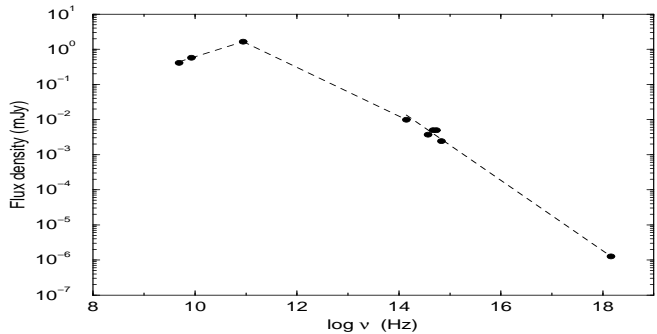,width=12cm, height=4.7cm, angle=0.0}  
    \caption{The multiwavelength spectrum of GRB 970508, on May 22, 1997.
           Adapted from Gorosabel (1999). See also Wijers and Galama (1998) 
           who estimated $\nu_{a}$ $\sim$ 2.5 x 10$^{9}$ Hz, 
           $\nu_{m}$ $\sim$ 10$^{11}$ Hz and $\nu_{c}$ $\sim$ 10$^{14}$ Hz.}
    \end{figure}

How does the GRB take place? The fireball model is based on the existence
of a ``central engine''. However, current
observations have been unable to give any insight on this issue. Until 1994, 
the number 
of theoretical models amounted up to 150 (see Nemiroff 1994), but once the
cosmological nature of GRBs was established, only those models  
dealing with mechanisms able to produce  10$^{51-54}$ erg in a 
few seconds were taken into consideration. They can be classified into two 
groups:

i) coalescence of neutron stars in a binary system
(Narayan et al. 1992): lifes of such systems are of the order of
$\sim$ 10$^{9}$ years, and large escape velocities are usual, putting them 
far away from the regions where their progenitors were born. The likely 
result is a Kerr black hole (BH), and the energy released energy during 
the merger process is $\sim$ 10$^{54}$ erg. It is also possible that a
$\sim$ 0.1 $M_{\odot}$ accretion disk forms around the BH and is accreted
within a few dozen seconds, then producing internal shocks leading to the
GRB (Katz 1997).  There are variations of these models where one or
two components are substituted for black holes (Paczy\'nski 1991).

ii) a ``failed'' type-I SN (Bodenhaimer and
Woosley 1983, Woosley 1993) or {\it hypernova} (as it has been called by 
Paczy\'nski (1998) on the basis of the observational consequences): 
very massive stars (Wolf-Rayet) collapse forming a Keck BH and a 
0.1-1 $M_{\odot}$ torus. 
The matter is accreted at a very high rate and the energy is
extracted via the rotational energy of the BH or via the accretion
energy from the disk. In any case, a ``dirty fireball'', is produced 
reaching a luminosity $\sim$ 300 times larger that than of a normal SN. 
This would happen every $\sim$ 10$^{6}$ yr. In this scenario, GRBs
would be produced in dense enviroments near star forming 
regions (see also MacFadyen and Woosley 1999) and GRBs might be used
for deriving the SFR in the Universe (Totani et al. 1998, Krumholz et al. 
1998).

\section{Summary}
The existence of X-ray afterglow in {\it most} bursts is confirmed.
Out ot 15 {\it BSAX} pointings, 11 revealed a strong afterglow, and 
4 displayed faint ones, leading to the detection of
optical/IR/radio counterparts in 1997-98. Very promising seems to be 
the determination of $z$ for the host galaxies by means of 
absorption edges or emission lines 
in the X-ray afterglow. This requires prompt X-ray follow-up, as
was achieved for GRB 970508 and GRB 970828.
However, only the population of bursts
with durations of few seconds has been explored. Short bursts lasting
less than 1 s, like GRB 980706, that follow the -3/2 slope in the 
log $N$-log $S$ diagram (in contrast to the longer bursts) remain to be 
detected at longer wavelengths.

It is generally believed is that internal shocks produced by the 
$^{\prime\prime}$central engine$^{\prime\prime}$ are responsible
for the $\gamma$-ray emission, whereas the slowing down of a relativistic 
shell on the interstellar medium (an external shock) would cause
the post-GRB emission at longer wavelengths (the afterglow). Energy
releases of $\sim$ 10$^{54}$ erg (as derived for GRB 980329) are difficult 
to reconcile with theoretical models and non-isotropic emission, such as 
intrinsic beaming appears as the most plaussible resolution of this problem.
 
{\it BSAX} and {\it RXTE} have opened a new window in the GRB field and 
it is widely accepted now that most GRBs, if not all, lie at cosmological 
distances. It is expected that {\it BSAX}, {\it RXTE} and {\it CGRO} 
will facilitate the discoveries of many other counterparts and, together with 
the new high-energy observatories 
({\it AXAF}, {\it SPECTRUM X/$\Gamma$}, {\it XMM}, {\it INTEGRAL}, 
{\it HETE 2}) and, hopefully {\it SWIFT} and {\it BALLERINA}, will 
help to solve the long-standing Gamma-Ray Burst mystery.



\begin{thebibliography}{}




\bibitem[\protect
\astroncite{Antonelli et al.}{1997}]{ant97}
Antonelli, L. al. 1997, IAU Circ. 6792

\bibitem[\protect
\astroncite{Atteia et al.}{1987}]{att87}
Atteia, J.-L. et al. 1987, ApJS 64, 305

\bibitem[\protect
\astroncite{Bartolini et~al.}{1998}]{bar98}
Bartolini, C. et al. 1998, in Gamma-Ray Bursts, eds. C. A. Meegan, R. Preece
and T. Koshut, AIP 428, 540.

\bibitem[\protect
\astroncite{Bloom et al.}{1998a}]{blo98a}
Bloom, J. S. et al. 1998a, ApJ 507, L25

\bibitem[\protect
\astroncite{Bloom et al.}{1998b}]{blo98b}
Bloom, J. S. et al. 1998b, ApJ 508, L21

\bibitem[\protect
\astroncite{Bloom and Kulkarni}{1998}]{bak98}
Bloom, J. S. and Kulkarni, S. R. 1998, GCN Circ. 161

\bibitem[\protect
\astroncite{Bodenhaimer and Woosley}{1983}]{bod83}
Bodenhaimer, P. and Woosley, S. 1983, ApJ 269, 281

\bibitem[\protect
\astroncite{Bond}{1997}]{bon97}
Bond, H. 1997, IAU Circ. 6655

\bibitem[\protect
\astroncite{Bonnell and Klebesadel}{1996}]{bon96}
Bonnell, J. T. and Klebesadel, R. W. 1996, in Gamma-Ray Bursts, eds. C. 
Kouveliotou, M. F.  Briggs and G. J. Fishman, AIP 384, 977

\bibitem[\protect
\astroncite{Bremer et al.}{1998}]{bre98}
Bremer, M. et al. 1998, A\&A 332, L13

\bibitem[\protect
\astroncite{Castander et al.}{1997}]{cas97}
Castander, F. et al. 1997, IAU Circ. 6791

\bibitem[\protect
\astroncite{Castro-Tirado et~al.}{1997}]{cas97}
Castro-Tirado, A. J. et al. 1997, IAU Circ. 6598



\bibitem[\protect
\astroncite{Castro-Tirado}{1998}]{cas98}
Castro-Tirado, A. J. 1998, in {\it Ultraviolet Astrophysics: beyond the IUE
Final Archive}, Sevilla, Spain, November 1997, eds: Gonzalez-Riestra, R., 
Wamsteker, W., Harris, R., ESA Conference Proceedings SP-413, pp. 659-668.
(http://xxx.lanl.gov/abs/astro-ph/9803007)


\bibitem[\protect
\astroncite{Castro-Tirado et al.}{1998a}]{cas98a}
Castro-Tirado, A. J. et al. 1998a, Sci 279, 1011

\bibitem[\protect
\astroncite{Castro-Tirado et al.}{1998b}]{cas98b}
Castro-Tirado, A. J. et al. 1998b, IAU Circ. 6848

\bibitem[\protect
\astroncite{Castro-Tirado et al.}{1999}]{cas99}
Castro-Tirado, A. J. et al. 1999, ApJ 
511, L85

\bibitem[\protect
\astroncite{Celidonio et al.}{1998}]{cel98}
Celidonio, G. et al. 1998, IAU Circ. 6851


\bibitem[\protect
\astroncite{Costa et al.}{1997a}]{cos97a}
Costa, E. et al. 1997, IAU Circ. 6533

\bibitem[\protect
\astroncite{Costa et al.}{1997b}]{cos97b}
Costa, E. et al. 1997b, Nat 387, 783

\bibitem[\protect
\astroncite{DiCiolo et al.}{1998}]{cio98}
DiCiolo, L. et al. 1998, IAU Circ. 7074

\bibitem[\protect
\astroncite{Diercks et al.}{1997}]{die97}
Diercks, A. et al. 1997, IAU Circ. 6791

\bibitem[\protect
\astroncite{Djorgovski et al.}{1997}]{djo97}
Djorgovski, S. G. et al. 1997, Nat 387, 876

\bibitem[\protect
\astroncite{Djorgovski et al.}{1998}]{djo98}
Djorgovski, S. G. et al. 1998, ApJ 508, L17

\bibitem[\protect
\astroncite{Djorgovski et al.}{1999}]{djo99}
Djorgovski, S. G. et al. 1999, GCN Circ. 189

\bibitem[\protect
\astroncite{Feroci et al.}{1998a}]{fer98a}
Feroci, M. et al. 1998a, A\&A 332, L29

\bibitem[\protect
\astroncite{Feroci et al.}{1998b}]{fer98b}
Feroci, M. et al. 1998b, IAU Circ. 6909

\bibitem[\protect
{Fishman and Meegan}{1995}]{fis95} 
Fishman, G. J. and Meegan, C. A. 1995, ARA\&A 33, 415

\bibitem[\protect
\astroncite{Frail et al.}{1997}]{fra97}
Frail, D., Kulkarni, Nicastro, L., Feroci, M. and Taylor, G. 1997, 
Nat 389, 261

\bibitem[\protect
\astroncite{Frail et al.}{1998}]{fra98}
Frail, D. et al. 1998, GCN Circ. 128

\bibitem[\protect
\astroncite{Frail et al.}{1999}]{fra99}
Frail, D. et al. 1999, in Proceedings of the {\it Gamma-ray Bursts in 
the Afterglow Era} Workshop, Rome, Nov 1998, A\&AS, in press.

\bibitem[\protect
\astroncite{Frontera et al.}{1998}]{fro98}
Frontera, F., Costa, E., Piro, L. et al. 1998, ApJ 493, L67

\bibitem[\protect
\astroncite{Frontera et al.}{1998b}]{fro98b}
Frontera, F. et al. 1998b, IAU Circ. 6853

\bibitem[\protect
\astroncite{Fruchter et al.}{1997}]{fru97}
Fruchter, A. et al. 1997, IAU Circ. 6747

\bibitem[\protect
\astroncite{Fruchter}{1999}]{fru99}
Fruchter, A. 1999, ApJ 512, L1

\bibitem[\protect
\astroncite{Fruchter et al.}{1999}]{fru99}
Fruchter, A. et al. 1999, ApJ, in press 
(http://xxx.lanl.gov/abs/astro-ph/9807295)

\bibitem[\protect
\astroncite{Galama et al.}{1997}]{gal97}
Galama, T. et al. 1997, Nat 387, 479

\bibitem[\protect
\astroncite{Galama et al.}{1998a}]{gal98a}
Galama, T. et al. 1998a, ApJ 497, L13

\bibitem[\protect
\astroncite{Galama et al.}{1998b}]{gal98b}
Galama, T. et al. 1998b, Nat 395, 670


\bibitem[\protect
\astroncite{Ginzburg and Syrovatskii}{1965}]{vit65}
Ginzburg, V. L. \& Syrovatskii, S. I. 1965, ARA\&A 3, 297

\bibitem[\protect
\astroncite{Goodman}{1986}]{god86}
Goodman, J. 1986, ApJ 308, L47

\bibitem[\protect
\astroncite{Gorosabel et al.}{1998a}]{gor98a}
Gorosabel, J. et al. 1998a, A\&A 335, L5

\bibitem[\protect
\astroncite{Gorosabel et al.}{1998b}]{gor98b}
Gorosabel, J. et al. 1998b, A\&A 339, 719

\bibitem[\protect
\astroncite{Gorosabel}{1999}]{gor99}
Gorosabel, J. 1999, Ph.D. Thesis, Universitat de Val\`encia, Spain

\bibitem[\protect
\astroncite{Graziani et al.}{1999}]{gra99}
Graziani, C. et al. 1999, in Proceedings of the {\it Gamma-ray Bursts in 
the Afterglow Era} Workshop, Rome, Nov 1998, A\&AS, in press

\bibitem[\protect
\astroncite{Greiner and Heise}{1997}]{gre97}
Greiner, J. \& Heise, J. 1997, IAU Circ. 6570


\bibitem[\protect
\astroncite{Greiner et al.}{1997}]{gre97}
Greiner, J., Englhauser, J., Groot, P. J. and Galama, T.
1997, IAU Circ. 6757

\bibitem[\protect
\astroncite{Greiner and Heise}{1997}]{gre97}
Greiner, J. \& Heise, J. 1997, IAU Circ. 6570

\bibitem[\protect
\astroncite{Greiner}{1999}]{gre99}
Greiner, J. 1999, http://www.obs.aip.de/~jcg/grbgen.html

\bibitem[\protect
\astroncite{Groot et al.}{1997}]{gro97}
Groot P.J. et al. 1997, IAU Circ. 6584 

\bibitem[\protect
\astroncite{Groot et al.}{1998}]{gro98}
Groot P.J. et al. 1998, ApJ 493, L27

\bibitem[\protect
\astroncite{Guarnieri et al.}{1997}]{gua97}
Guarnieri, A. et al. 1997, A\&A 328, L13

\bibitem[\protect
\astroncite{Halpern et al.}{1997}]{hal97}
Halpern, J. et al. 1997, IAU Circ. 6788

\bibitem[\protect
\astroncite{Hjorth et al.}{1999}]{hjo99}
Hjorth, J. et al. 1999,  in Proceedings of the {\it Gamma-ray Bursts in 
the Afterglow Era} Workshop, Rome, Nov 1998, A\&AS, in press.

\bibitem[\protect
\astroncite{Hurley et al.}{1999}]{hur99}
Hurley, K. et al. 1999, GCN Circ. 129

\bibitem[\protect
\astroncite{Funk et al.}{1996}]{fun96}
Funk, B. et al. 1996, in Gamma-Ray Bursts, eds. C. Kouveliotou, M. F.



\bibitem[\protect
\astroncite{in\' \rm t Zand et al.}{1998}]{zan98}
in\' \rm t Zand, J. et al. 1998, IAU Circ. 6805

\bibitem[\protect
\astroncite{Katz}{1997}]{kat97}
Katz, J. I. 1997, ApJ 490, 633

\bibitem[\protect
\astroncite{Klebesadel, Olsen and Strong}{1973}]{keb73}
Klebesadel, R., Olsen, R. and Strong, I. 1973, ApJ 182, L85

\bibitem[\protect
\astroncite{Klose et al.}{1998}]{klo98}
Klose, S., Meusinger, H. and Lehmann, H. 1998, IAUC 6864

\bibitem[\protect
\astroncite{Krumholz et~al.}{1998}]{kru98}
Krumholz, M., Thorsett, S. and Harrison, F. 1998, ApJ 506, L81 

\bibitem[\protect
\astroncite{Kulkarni et al.}{1998a}]{kul98a}
Kulkarni, S. et al. 1998, Nat 393, 35

\bibitem[\protect
\astroncite{Kulkarni et al.}{1998b}]{kul98b}
Kulkarni, S. et al. 1998, Nat 395, 663

\bibitem[\protect
\astroncite{Levine et al.}{1998}]{lev98}
Levine, A. et al. 1998, IAU Circ. 6966

\bibitem[\protect
\astroncite{MacFadyen and Woosley}{1999}]{mac99}
MacFadyen, A. and Woosley, S. 1999, ApJ, in press 
(http://xxx.lanl.gov/abs/astro-ph/9810274)

\bibitem[\protect
\astroncite{Marshall et al.}{1997}]{mar97}
Marshall. F. et al. 1997, IAU Circ. 6683


\bibitem[\protect
\astroncite{Mazets et al.}{1981}]{maz81}
Mazets, E. P. et al. 1981, Ap\&SS 80, 3

\bibitem[\protect
\astroncite{Meegan, C. et al.}{1992}]{mee92}
Meegan, C. A. et al. 1992, Nat 355, 143


\bibitem[\protect
\astroncite{M\'esz\'aros et al.}{1994}]{mes94}
M\'esz\'aros, P. et al. 1994, ApJ 432, 181


\bibitem[\protect
\astroncite{M\'esz\'aros and Rees}{1997}]{mes97}
M\'esz\'aros, P. and Rees, M. J. 1997, ApJ 476, 232

\bibitem[\protect
\astroncite{M\'esz\'aros and Rees}{1998}]{mes98}
M\'esz\'aros, P. and Rees, M. J. 1998, MNRAS 299, L10


\bibitem[\protect
\astroncite{Metzger et al.}{1997b}]{met97}
Metzger, M. R. et al. 1997, Nat 387, 878


\bibitem[\protect
\astroncite{Muller et al.}{1998}]{mul98}
Muller, J. et al. 1998, IAU Circ. 6910

\bibitem[\protect
\astroncite{Murakami et~al.}{1997}]{mur97}
Murakami, T. et al. 1996, IAU Circ. 6732

\bibitem[\protect
\astroncite{Narayan et al.}{1992}]{nar92}
Narayan, R. et al. 1992, ApJ 395, L83 

\bibitem[\protect
\astroncite{Nemiroff}{1994}]{nem94}
Nemiroff, R. 1994, Comments Astrophys. 17, 189

\bibitem[\protect
\astroncite{Palazzi et al.}{1998}]{pal98}
Palazzi, E. et al. 1998, A\&A 336, L95 

\bibitem[\protect
\astroncite{Paczy\'nski}{1986}]{pac86}
Paczy\'nski, B. 1986, ApJ 308, L43

\bibitem[\protect
\astroncite{Paczy\'nski}{1986}]{pac86}
Paczy\'nski, B. 1991, Acta Astronomica 41, 257

\bibitem[\protect
\astroncite{Paczy\'nski}{1998}]{pac98}
Paczy\'nski, B. 1998, ApJ 494, L45

\bibitem[\protect
\astroncite{Pedersen et~al.}{1998}]{ped98}
Pedersen, H. et al. 1998, ApJ 496, 311

\bibitem[\protect
\astroncite{Pedichini et~al.}{1997}]{ped97}
Pedichini, F. et al. 1998, A\&A 327, L36

\bibitem[\protect
\astroncite{Pian et~al.}{1998a}]{pia98a}
Pian, E. et al. 1998a, ApJ 492, L103

\bibitem[\protect
\astroncite{Pian et~al.}{1998b}]{pia98b}
Pian, E. et al. 1998b, in Proceedings of the {\it Gamma-ray Bursts in 
the Afterglow Era} Workshop, Rome, Nov 1998, A\&AS, in press 
(http://xxx.lanl.gov/abs/astro-ph/9903113)

\bibitem[\protect
\astroncite{Piran et al.}{1999}]{pir99}
Piran, T. et al. 1999, Physics Reports, in press (http://xxx.lanl.gov/abs/astro-ph/astro-ph/9810256)

\bibitem[\protect
\astroncite{Piro et al.}{1996}]{pir96}
Piro, L. et al. 1996, IAU Circ. 6467

\bibitem[\protect
\astroncite{Piro et al.}{1997a}]{pir97a}
Piro, L. et al. 1997a, IAU Circ. 6570

\bibitem[\protect
\astroncite{Piro et~al.}{1997a}]{pir97b}
Piro, L. et al. 1997b, IAU Circ. 6617

\bibitem[\protect
\astroncite{Piro et~al.}{1997c}]{pir97c}
Piro, L. et al. 1997c, IAU Circ. 6656

\bibitem[\protect
\astroncite{Piro et~al.}{1998}]{pir98}
Piro, L. et al. 1998, GCN Circ. 155

\bibitem[\protect
\astroncite{Piro et~al.}{1999}]{pir98}
Piro, L. et al. 1999, ApJ, in press (astro-ph/9902013)

\bibitem[\protect
\astroncite{Pooley and Green}{1997}]{poo97}
Pooley, G. and Green, D. 1997, IAU Circ. 6670 

\bibitem[\protect
\astroncite{Remillard et~al.}{1997}]{rem97}
Remillard, R., Wood, A., Smith, D. and Levine, A. 1997 IAU Circ. 6726

\bibitem[\protect
\astroncite{Sahu et~al.}{1997}]{sah97}
Sahu, K. et al. 1997, Nat 387, 476 

\bibitem[\protect
\astroncite{Sari et al.}{1998}]{sari98}
Sari, R., Piran, T. and Narayan, R. 1998, ApJ 497, L17


\bibitem[\protect
\astroncite{Smith et al.}{1997}]{smi97}
Smith, D. A. et al. 1997, IAU Circ. 6718

\bibitem[\protect
\astroncite{Smith et al.}{1998}]{smi98a}
Smith, M. J. S. et al. 1998a, IAU Circ. 6938

\bibitem[\protect
\astroncite{Smith et al.}{1998}]{smi98b}
Smith, D. A. et al. 1998b, IAU Circ. 6938

\bibitem[\protect
\astroncite{Smith et al.}{1999}]{smi99}
Smith, I. A. et al. 1999, A\&A, in press
(http://xxx.lanl.gov/abs/astro-ph/9811026)

\bibitem[\protect
\astroncite{Sofitta et~al.}{1997}]{sof97}
Sofitta, P. et al. 1997, IAU Circ. 6797

\bibitem[\protect
\astroncite{Sofitta et~al.}{1998}]{sof98}
Sofitta, P. et al. 1998, IAU Circ. 6884

\bibitem[\protect
\astroncite{Sokolov et al.}{1998}]{sok98}
Sokolov, V. V. et al. 1998a, A\&A 334, 117


\bibitem[\protect
\astroncite{Tavani}{1997}]{tav97}
Tavani, M. 1997, ApJ 483, L87

\bibitem[\protect
\astroncite{Tanvir}{1997}]{tan97}
Tanvir, N. et al. 1997, IAU Circ. 6796

\bibitem[\protect
\astroncite{Taylor}{1997}]{tay97}
Taylor, G. B. et al. 1997, Nat 389, 263

\bibitem[\protect
\astroncite{Taylor}{1998}]{tay98}
Taylor, G. B. et al. 1998, ApJ 502, L115

\bibitem[\protect
\astroncite{Totani et~al.}{1998}]{tot98}
Totani, T. 1998, ApJ, in press 
(http://xxx.lanl.gov/abs/astro-ph/9805263)

\bibitem[\protect
\astroncite{van Paradijs et~al.}{1997}]{par97}
van Paradijs, J. et al. 1997. Nat 386, 686

\bibitem[\protect
\astroncite{Vietri}{1997}]{vie97}
Vietri, M. 1997, ApJ 488, L105

\bibitem[\protect
\astroncite{Waxman}{1997}]{wax97}
Waxman, E. 1997, ApJ 489, L33

\bibitem[\protect
\astroncite{Wijers et~al.}{1997}]{wij97}
Wijers, R. A. M. J., Rees, M. J. and M\'esz\'aros, P. 1997, MNRAS 288, L51

\bibitem[\protect
\astroncite{Wijers and Galama}{1998}]{wij98}
Wijers, R. A. M. J. and Galama, T. J. 1998, ApJ, in press 
(http://xxx.lanl.gov/abs/astro-ph/9805341)

\bibitem[\protect
\astroncite{Woosley}{1993}]{woo73}
Woosley, S. 1993, ApJ 405, 273

\bibitem[\protect
\astroncite{Yoshida et al.}{1999}]{yos99}
Yoshida, A. et al. 1999, in Proceedings of the {\it Gamma-ray Bursts in 
the Afterglow Era} Workshop, Rome, Nov 1998, A\&AS, in press.

\bibitem[\protect
\astroncite{Zapatero-Osorio et al.}{1998}]{zap98}
Zapatero-Osorio, M. R. et al. 1998, IAU Circ. 6747

\bibitem[\protect
\astroncite{Zharikov et al.}{1998}]{zar98}
Zharikov, S. V., Sokolov, V. V., Barishev, Yu. V. 1998,  A\&A 337, 356 

\end{thebibliography}
\end{document}